\documentclass[%
 reprint,
superscriptaddress,
longbibliography,
 amsmath,amssymb,
 prx,
]{revtex4-2}

\usepackage{graphicx}
\graphicspath{ {./images/} }
\usepackage[dvipsnames]{xcolor}
\usepackage{dcolumn}
\usepackage{bm}
\usepackage{physics}
\usepackage[colorlinks=true,citecolor=blue,linkcolor=red,linktocpage=true]{hyperref}
\usepackage{bbm} 
\usepackage{makecell} 

\begin{document}

\preprint{APS/123-QED}

\title{Non-reciprocity is necessary for robust dimensional reduction and strong responses in stochastic topological systems
}

\author{Aleksandra Nelson}
 \affiliation{CTBP, Rice University, Houston TX, USA}
\author{Evelyn Tang}%
\affiliation{CTBP, Rice University, Houston TX, USA}
\affiliation{%
 Physics department, Rice University, Houston TX, USA
}


\begin{abstract}

 Topological theory predicts the necessary conditions for robust dimensional reduction in a host of quantum and classical systems. Models have recently been  proposed for stochastic systems which describe many biological and chemical phenomena. However, general theoretical principles are lacking for this class of systems, exemplified by the breakdown of the celebrated bulk-edge correspondence. We prove that contrary to topological phases in quantum and many classical systems, stochastic systems require non-reciprocal (or non-Hermitian) transitions for robust edge responses, which holds across all dimensions and geometries. We propose a novel explanation of hybridization that destroys edge responses in reciprocal (Hermitian) stochastic systems. Further, we show that stochastic steady states grow dramatically with non-reciprocity, in contrast to their quantum counterparts which plateaus. We analyze the resulting theoretical
and physical consequences and how non-reciprocity mitigates the effects of hybridization towards robust edge states in equilibrium and non-equilibrium steady states. These results establishes the crucial role of non-reciprocal interactions  for responses that are robust to random perturbations in active and living matter.

\end{abstract}

\maketitle

\section{Introduction}

{Topological invariants have proved useful for analyzing emergent function as they characterize a property of the entire system, and are insensitive to local details, disorder, and noise \cite{hasan2010, qi2011, bansil2016}. They support edge states, which reduce the system response to a lower dimensional space and offer a mechanism for the emergence of localization or global cycles within a large space of reactions.} Topological phases have been realized in a variety of platforms, such as quantum electronic systems \cite{hasan2010, qi2011, bansil2016}, photonic crystals \cite{lu2014, ozawa2019, tang2022}, acoustic crystals \cite{ma2019, xin2020, zheng2022}, mechanical lattices \cite{mao2018, xin2020}, electrical circuits \cite{lee2018, imhof2018, hofmann2019}, {chiral fluids \cite{dasbiswas2018, souslov2019}, and }active matter \cite{ghatak2020, sone2020, palacios2021, shankar2022}. 

{The broad generality of topology stems from its origins in random matrix theory \cite{mehta1967, schnyder2008, ryu2010}, that provides the necessary and sufficient conditions for targeted steady state responses that are also robust to random perturbations \cite{hasan2010, moore2010}. This allows for a systematic study of the gapless or floppy modes \cite{kane2014, chen2014, paulose2015} that remain accessible under arbitrary perturbations of the system, including disorder.  All possible phases can be classified based only on the system's symmetry, providing an exhaustive list for the existence of models with robust edge responses \cite{schnyder2008, ryu2010}. This has led to several key theoretical predictions in quantum electronic systems \cite{hasan2010, qi2011, tang2011, tang2014} that were subsequently verified experimentally \cite{ye2018, mao2020}. New physics has similarly been predicted and observed in other fields, such as dissipationless signal propagation in photonic crystals \cite{haldane2008, wang2008, wang2009}.} 

Due to their attractive properties, it would be desirable to realize these phases in a broader class of {accessible} platforms. Promising realizations of these topological invariants in stochastic systems have been proposed in biology and robotics. Edge currents have been used to model long and emergent oscillations, such as the circadian rhythm in cyanobacteria \cite{zheng2024} -- showing high coherence and efficiency with unusually few free parameters and novel experimental signatures. Edge localization has been shown to support high sensitivity for adapting organisms \cite{murugan2017}, e.g., when \textit{E. coli} perform chemotaxis. In synthetic systems, a topological mechanism would allow a microswimmer to maximize its greatest available phase space and hence swimming speed \cite{tang2021}. Crucially, these topological mechanisms can prove robust to environmental changes or system malfunctions, within a flexible design architecture.

However, our understanding of stochastic topological phases is based on isolated examples \cite{murugan2017, dasbiswas2018, knebel2020, yoshida2021, tang2021, sawada2024a, zheng2024} with little general theory thus far. Specifically, despite the presence of edge states and currents in previous works, it is unclear how and under which conditions topological invariants in stochastic systems lead to robust edge responses. This lack of general theory prevents a deeper understanding or  application of stochastic topological models despite the advantage of robustness. 
Meanwhile, edge responses have been well-studied in quantum systems, where they are caused by bulk topological invariants -- a principle known as  the bulk-edge correspondence \cite{haldane1988, kane2005a, rhim2018}. 
This gap {in our understanding} demands a systematic comparison between quantum and stochastic topological systems, in order to know whether stochastic topological phases are governed by similar principles and criteria as quantum topological phases. More broadly, the distinction between quantum and stochastic operators on the same network is precisely the distinction between adjacency and Laplacian matrices -- both of which are widely used in graph theory and dynamical systems \cite{newman2018networks}. Hence, knowledge of the differences between specific quantum and stochastic topological models is relevant for characterizing many other systems and their dynamics \cite{wong2016,mirzaev2013}.

A key result of this work is that contrary to quantum systems, stochastic topological systems must be non-reciprocal to have an edge response.
As stochastic transitions are real-valued, non-reciprocity in stochastic systems is equivalent to non-Hermiticity, which has been of great recent interest. Non-Hermitian physics is used to describe driven and active systems, both quantum and classical \cite{ashida2020, hosaka2022, shankar2022}.
{Notably, non-Hermiticity introduces new aspects to topological classification \cite{kawabata2019}. For instance, the complex spectrum of non-Hermitian systems leads to new topological invariants expressed in terms of eigenvalue winding \cite{ghatak2019}, which have no Hermitian counterparts. These new phenomena lead to unique behavior in non-Hermitian systems. Examples include the skin effect \cite{yao2018, borgnia2020, okuma2020, zhang2020, zhang2022a}, where sensitivity to boundary conditions leads to an extensive number of edge states, or exceptional points \cite{kawabata2019a, ding2022, heiss2012}, which are known to protect edge modes in some quantum and classical systems \cite{sone2020}.}
Our work highlights the crucial role of non-reciprocity, or non-Hermiticity, for edge responses in stochastic topological systems and, hence, for robust behavior in soft and living matter.

To explore these questions, we develop new analyses to compare quantum and stochastic systems defined on the same network. We demonstrate for reciprocal stochastic systems the breakdown of key features in other established topological systems, such as the bulk-boundary correspondence. We   introduce the mechanism of hybridization as a novel explanation to account for this. Next, the effects of non-reciprocity are investigated in a one-dimensional model with point gap topology and a two-dimensional model exhibiting steady state currents. We find that the quantum response depends very weakly on non-reciprocity, while the stochastic response grows dramatically with non-reciprocity. Thus, non-reciprocity mitigates the effects of hybridization towards robust edge responses in stochastic systems. Having established the necessity and importance of non-reciprocity for topological stochastic systems, we then analyze the many theoretical and physical consequences of this result. Lastly, we demonstrate a novel mechanism of reducing hybridization to generate edge responses in an understudied class of non-Hermitian systems.

\begin{figure}
    \centering
    \includegraphics{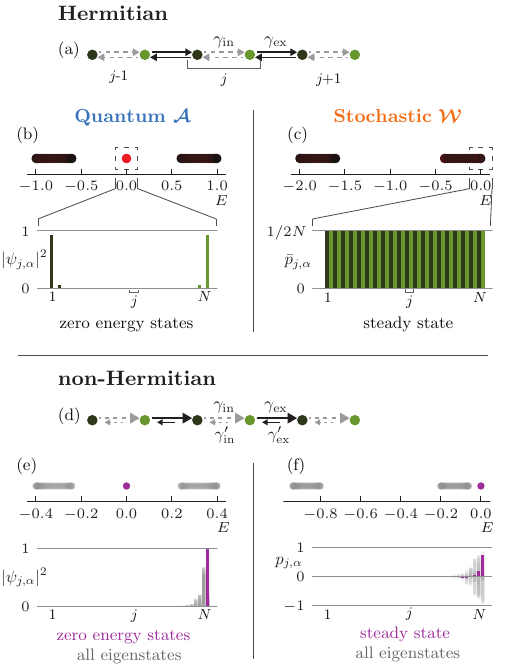}
    \caption{\textbf{Non-Hermiticity is necessary for topological edge states in stochastic systems.}
    We contrast the behavior of Hermitian (top half) and non-Hermitian (bottom half) Su-Schriefer-Heeger (SSH) models \cite{su1980, yao2018} describing quantum (left column) and stochastic (right column) systems. 
    (a) Hermitian SSH model has reciprocal transition rates.
    (b) 
    {Top: the spectrum of the quantum Hermitian topological model has two bulk bands (black) and two edge states at $E=0$ (red).
    Bottom: probability of the zero energy states show localization at the network edges.}
    (c) 
    Top: spectrum of the stochastic Hermitian model with the same topological invariant has only bulk states (black). Edge states are hybridized with the bulk states and lose their localization.
    Bottom: the steady state probability is uniform across the network.
    (d) Transition rates in the non-Hermitian SSH model are non-reciprocal.
    (e, f) Both quantum and stochastic non-Hermitian models have all eigenstates localized on the right edge, {including the $E=0$ modes in purple and all other states in gray}.
    All plots are made with open boundary conditions; panels (b, c) are plotted for $\gamma_\textnormal{in}=0.2$, $\gamma_\textnormal{ex}=0.8$, $N=20$, and panels (e, f) for $\gamma_\textnormal{ex}=0.64$, $\gamma_\textnormal{ex}'=\gamma_\textnormal{in}=0.16$, $\gamma_\textnormal{in}'=0.04$.
    }
    \label{fig:SSH_spectrum_states}
\end{figure}

\section{Non-Hermiticity is necessary for topological edge states in stochastic systems}

\label{sec:NH_necessary}

{We study stochastic systems described by a master equation $\partial_t \bm{p}=\mathcal{W}\bm{p}$, where the vector $\bm{p}$ describes the probability at each state. The transition matrix $\mathcal{W}$ has two parts $\mathcal{W}=\mathcal{A}-\mathcal{D}$, where $\mathcal{A}_{ij}=\bra{i}\ket{j}$ describes the transition rates between different states $i\neq j$ while the diagonal matrix $\mathcal{D}_{ij}=\delta_{ij}\sum_{k\neq j}\mathcal{A}_{kj}$ is the sum of the rates leaving each state \cite{schnakenberg1976}. This term $\mathcal{D}$ enforces probability conservation by making $\sum_{i}\mathcal{W}_{ij}=0$. 
}

{The two parts of $\mathcal{W}$ naturally allow us to compare quantum and stochastic systems that operate on the same network, since the off-diagonal matrix $\mathcal{A}$ has a similar form to a tight-binding Hamiltonian typically used for electronic systems \cite{ashcroft1976}. Hence, the quantum and stochastic descriptions share the same transitions between different states, specified by the matrix $\mathcal{A}$. The difference between both systems is only due to the diagonal term $\mathcal{D}$ that conserves probability.} {To illustrate this, we consider quantum and stochastic systems on the well-studied 1D Su-Schrieffer-Heeger (SSH) model \cite{su1980, asboth2016} (see Fig.~\ref{fig:SSH_spectrum_states}(a)). 
This 1D lattice has two states in each unit cell, i.e. $\ket{j,\mathrm{A}}$ and $\ket{j,\mathrm{B}}$ in the $j$-th cell, indicated by dark and light green circles respectively in Fig.~\ref{fig:SSH_spectrum_states}(a). There are two different transition rates: an internal rate $\gamma_\textnormal{in}$ within the unit cell, and an external rate $\gamma_\textnormal{ex}$ between unit cells. The Hamiltonian with open boundary conditions is
\begin{equation}
    \mathcal{A}=\left(\sum_{j=0}^N\gamma_\textnormal{in}\ket{j,\mathrm{B}}\bra{j,\mathrm{A}}+\sum_{j=0}^{N-1}\gamma_\textnormal{ex}\ket{j+1,\mathrm{A}}\bra{j,\mathrm{B}}\right)+\mathrm{tr.},
\end{equation}
where tr. stands for the transpose.} 

{We can now compare the behavior of the longest-lived eigenmodes in both the quantum and stochastic systems, described by matrices $\mathcal{A}$ and $\mathcal{W}$ respectively. In particular, we are interested in the topological phase, which occurs when $\gamma_\textnormal{ex}>\gamma_\textnormal{in}$, that is known to exhibit edge states in the quantum case \cite{asboth2016}. Indeed, the spectra $E$ of both systems look mostly similar (see Fig.~\ref{fig:SSH_spectrum_states}(b,c), top), except for two things. Firstly, there is a constant shift between them so that the quantum spectrum is symmetric around 0, while the stochastic spectrum has its maximum value at 0. The second difference is that there is a state within the gap in the quantum case (marked in red in Fig.~\ref{fig:SSH_spectrum_states}(b)) that is absent in the stochastic case. These differences mean that the longest-lived eigenmodes with $E=0$ come from different parts of the spectrum (highlighted in Figs.~\ref{fig:SSH_spectrum_states}(b) and (c) respectively with a dashed rectangle). In the quantum case, this zero-energy eigenstate $\bm{\psi}$ is clearly  localized on the system edge (see Fig.~\ref{fig:SSH_spectrum_states}(b)). In contrast, the steady state $\bm{\bar{p}}$ in the stochastic case is completely uniform (see Fig.~\ref{fig:SSH_spectrum_states}(c)).}

We can prove more generally that \textit{the stochastic steady state is always uniform for any ergodic Hermitian system}. By combining probability conservation $\sum_i\mathcal{W}_{ij}=0$ with Hermiticity $\mathcal{W}_{ij}=\mathcal{W}_{ji}$, we obtain that the sum of rates leaving state $j$ is equal to the sum of rates arriving to this state $-\mathcal{W}_{jj}=\sum_{i\neq j}\mathcal{W}_{ji}$. Now, the uniform state $\bar{p}_i=1/N$ will produce a zero net probability flow $\sum_i\mathcal{W}_{ji}\bar{p}_i=0$, i.e. the condition for the steady state. The existence of any other steady state is forbidden, since the steady state is unique in ergodic networks according to the Perron–Frobenius theorem \cite{perron1907, frobenius1912} applied to stochastic processes (Sec.~4 in Ref.~\cite{fernengel2022}). Thus, the steady state is uniform in any ergodic reciprocal network. Note that ergodic networks are connected: any state $j$ can be reached from any other state $i$ by a series of transitions -- which is the case in typical tight-binding and topological models. Remarkably, no restrictions besides reciprocity and ergodicity were made about the type of stochastic system that this proof holds for, so it applies to any geometry and dimension.

{This lack of edge states in the stochastic system is surprising as edge states are generally protected by the network topology. More specifically, a non-zero topological invariant calculated under periodic boundary conditions gives rise to edge states under open boundary conditions, according to the bulk-edge correspondence \cite{haldane1988, kane2005a, rhim2018}. In this case, the Zak phase invariant is calculated using eigenvectors of the quantum Hamiltonian $\mathcal{H}$ (\cite{berry1983, zak1989}; also see Appendix \ref{app:SSH_topology}), and is actually the same for the stochastic system.} {After all, the master equation is similar to the Schr\"{o}dinger equation for quantum dynamics $i\partial_t \psi=\mathcal{H}\psi$: observe that one operator can be mapped onto the other using $\mathcal{H}=i\mathcal{W}$ \cite{tang2021, sawada2024a}. Thus we can calculate topological invariants for stochastic systems in the same way as we would for a quantum system, since both operators have the same eigenvectors.  

In this Hermitian SSH model,
the matrix $\mathcal{D}$ is proportional to the identity matrix, so both $\mathcal{A}$ and $\mathcal{W}$ have the same eigenvectors and hence identical topological properties.
Yet, despite that we have the same Zak phase invariant in both quantum and stochastic models, the stochastic system does not have an edge state and breaks the bulk-edge correspondence. This is because under open boundary conditions, the matrix $\mathcal{D}$ is not proportional to the identity due to boundary effects, and hence alters the eigenvectors of stochastic matrix.
Effectively, matrix $\mathcal{D}$ shifts the edge states away from the bulk gap and hybridizes them with the bulk states, a novel explanation that we propose and will analyze in greater detail subsequently.

To obtain a non-uniform steady state, we need to introduce non-Hermitian transition rates. This is the non-Hermitian SSH model (Fig.~\ref{fig:SSH_spectrum_states}(d)), with backward rates $\gamma_\textnormal{in}'$ and $\gamma_\textnormal{ex}'$ in addition to each forward rate $\gamma_\textnormal{in}$ and $\gamma_\textnormal{ex}$.
\begin{align}
    \mathcal{A}
    &=\sum_{j=0}^N\gamma_\textnormal{in}\ket{j,\mathrm{B}}\bra{j,\mathrm{A}}+\gamma_\textnormal{in}'\ket{j,\mathrm{A}}\bra{j,\mathrm{B}}\nonumber\\
    &+\sum_{j=0}^{N-1}\gamma_\textnormal{ex}\ket{j+1,\mathrm{A}}\bra{j,\mathrm{B}}+\gamma_\textnormal{ex}'\ket{j,\mathrm{B}}\bra{j+1,\mathrm{A}}.
    \label{eq:NH_SSH}
\end{align}
To simplify this four-parameter space, we use two parameters that determine the physics. The ratio $r=(\gamma_\textnormal{ex}-\gamma_\textnormal{in})/(\gamma_\textnormal{ex}+\gamma_\textnormal{in})$ determines the balance between external and internal rates, where $r=0$ is the transition between the topological and trivial phases in the Hermitian limit. The non-reciprocity $c=(\gamma_\textnormal{in}-\gamma_\textnormal{in}')/(\gamma_\textnormal{in}+\gamma_\textnormal{in}')=(\gamma_\textnormal{ex}-\gamma_\textnormal{ex}')/(\gamma_\textnormal{ex}+\gamma_\textnormal{ex}')$  determines the balance between forward and backward rates, where $c=0$ is the Hermitian limit and $c=\pm 1$ is fully chiral or non-Hermitian.

When this model is non Hermitian $c\neq 0$, both the quantum and stochastic models have edge states at the $E=0$ mode (purple states in Fig.~\ref{fig:SSH_spectrum_states}(e, f)). Such localization in our 1D model is captured by a non-Hermitian topological invariant. This model has time-reversal symmetry: $\mathcal{T}_+\mathcal{A}^*\mathcal{T}_+^{-1}=\mathcal{A}$ and likewise for $\mathcal{W}$, which place them in the symmetry class AI \cite{kawabata2019}. This has a spectral  winding number invariant, which is non-zero when $c\neq 0$ (details in Appendix~\ref{app:SSH_NH}). Evidently, the bulk-edge correspondence is restored with non-Hermiticity.

Interestingly, besides the $E=0$ mode showing an edge state, an extensive number of other modes are also localized on the same edge (grey states in Fig.~\ref{fig:SSH_spectrum_states}(e, f)). This is the skin effect \cite{okuma2020}, a uniquely non-Hermitian phenomenon that occurs in both the quantum and stochastic systems. Extensivity of edge states in the presence of the skin effect removes the problem of hybridization induced by matrix $\mathcal{D}$. This is because if most of the states are localized on one edge, their hybridization still leaves localized edge states. Note that while our system appears to have topologically trivial values at only the single value of $c=0$, this is due to our simplified parameterization. Using the more general parametrization with four transition rates as in Eq.~\eqref{eq:NH_SSH}, one can obtain a trivial phase also in the non-reciprocal regime when $\gamma_\mathrm{ex}/\gamma_\mathrm{in}'=\gamma_\mathrm{ex}'/\gamma_\mathrm{in}$. Hence, our system has a finite range of values for both topological and trivial phases.

\begin{figure*}
    \centering
    \includegraphics{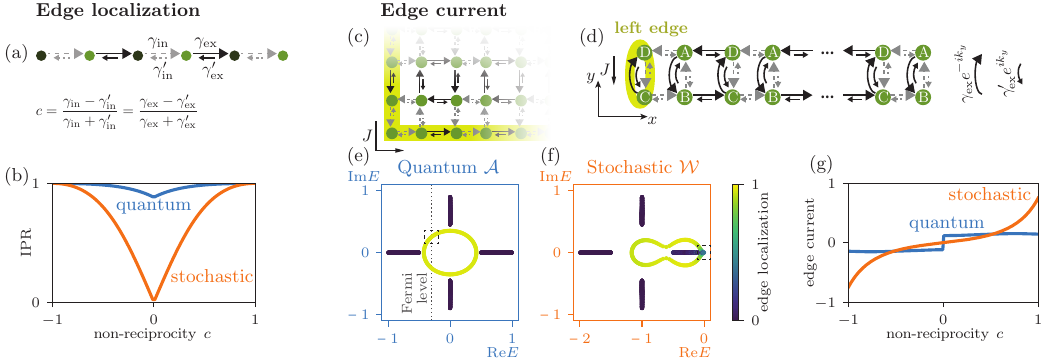}
    \caption{\textbf{Edge localization grows dramatically with non-reciprocity in stochastic systems but plateaus in quantum systems.}
    (a) Non-Hermitian SSH chain is parameterized by the ratio between external and internal rates $r=(\gamma_\textnormal{ex}-\gamma_\textnormal{in})/(\gamma_\textnormal{ex}+\gamma_\textnormal{in})$ and by non-reciprocity or chirality $c$; the latter controls the strength of non-Hermiticity.
    (b) The inverse participation ratio (IPR) quantifies localization of the stochastic (orange line) and quantum (blue line) {$E=0$ eigenstate: the quantum system increases slightly with non-reciprocity while the stochastic response is dramatic. }
    (c) This difference also holds for an edge current in a 2D non-Hermitian SSH model \cite{tang2021}.
    (d)~The fully finite network is approximated by a ribbon with open boundary condition along the $x$-axis and periodic boundary condition along the $y$-axis. We focus on currents on the left edge of the ribbon (highlighted with yellow). 
    (e) In the quantum spectrum, where shades of purple-to-yellow indicate degree of edge localization, the persistent current at $1/4$ filling is calculated in the state indicated with a dashed rectangle.
    (f) In the stochastic spectrum, the current is calculated in the steady state (indicated with a dashed rectangle), which is given by hybridization of the bulk and edge states. 
    (g) Again, the stochastic  edge current (orange) grows with non-reciprocity $c$, while the quantum current (blue) plateaus, just as in the 1D model.
    Panel (b) is plotted for $r=0.6$, panel (g) for $r=0.5$, panels (e, f) for $r=0.5$, $c=0.9$.}
    \label{fig:edge_chirality}
\end{figure*}

\section{Edge localization grows dramatically with non-reciprocity in stochastic systems but plateaus in quantum systems}
Having seen that non-Hermiticity is necessary for stochastic edge states, we ask whether these states are similar to edge states in non-Hermitian quantum systems.
\label{sec:non-recipr_growth}
To quantify a state localization we calculate its inverse participation ratio \cite{thouless1974} as 
$\mathrm{IPR}(\psi)=\sum_{j=1}^N|\psi_{j}|^4/\bra{\psi}\ket{\psi}^2$.
For the non-Hermitian SSH chain (Fig.~\ref{fig:edge_chirality}(a)), we calculate the IPR for the quantum zero-energy state and the stochastic steady state as a function of non-reciprocity $c$ (see Appendix~\ref{app:IPR} for derivation of IPR).
We observe that in the quantum system, localization depends weakly on non-reciprocity, while in the stochastic system, it grows with non-reciprocity and peaks in a fully non-reciprocal chain at $c=\pm1$ (Fig.~\ref{fig:edge_chirality}(b)).
To understand this behavior we note that in the Hermitian quantum model at $c=0$, there are edge states due to the Zak phase. Therefore, the non-Hermitian spectral winding \cite{yin2018} modifies localization only weakly.
In the stochastic case, we showed that there is no localization in the Hermitian limit and hence localization grows steadily as non-reciprocity increases.

{Still}, we would like to understand if this different dependence on non-reciprocity in stochastic and quantum edge states holds more generally, i.e., outside of states with a spectral winding {invariant}. Hence, we turn to a model with a qualitatively different physical response, that of edge currents. This is a 2D non-Hermitian SSH model \cite{tang2021, zheng2024}, {with the unit cell consisting of four states A, B, C and D (see Fig.~\ref{fig:edge_chirality}(c) and Appendix~\ref{app:2DSSH_matr} for details about the Hamiltonian). Transition rates in this model form two cycles, the internal cycle with forward rates $\gamma_\textnormal{in}$ that go counter-clockwise within a unit cell, and the external cycle with forward rates $\gamma_\textnormal{ex}$ that go clockwise between unit cells. Each forward rate has the corresponding back rate $\gamma_\textnormal{in}'$ and $\gamma_\textnormal{ex}'$. We adopt the same parametrization as in the 1D model: the ratio between external and internal rates $r$ and non-reciprocity $c$. 
Both the quantum and stochastic systems have inversion symmetry, which in Hermitian systems supports the Zak phase topological invariant \cite{liu2017, obana2019}. In the non-Hermitian versions, this topological invariant can be computed using biorthogonal eigenvectors \cite{lieu2018, zheng2024}. When the non-Hermitian Zak phase is calculated to be $(\Phi_x,\Phi_y)=(\pi,\pi)$, which occurs when the ratio between external and internal rates is $r>0$, our system has an edge current (highlighted with yellow in Fig.~\ref{fig:edge_chirality}(c)). When $r<0$ and the Zak phase is $(\Phi_x,\Phi_y)=(0, 0)$, there are no edge modes. More details about the Zak phase are provided in Appendix~\ref{app:2DSSH_topo}.}

To simplify the analysis, we approximate the model with all open boundaries by a ribbon with an open boundary condition in the $x$ direction and a periodic boundary condition in the $y$ direction, which guarantees a well-defined momentum $k_y$ (Fig.~\ref{fig:edge_chirality}(d)). As we show in Appendix~\ref{app:2DSSH_matr}, the fully open and the ribbon models have similar spectra for large enough systems.
We study currents flowing along the left edge of the ribbon, {highlighted with yellow in Fig.~\ref{fig:edge_chirality}(d).}

Quantum and stochastic persistent currents originate from different areas of the spectrum. This demands different methods to calculate the current in each case, which we outline below.
Quantum persistent current can only exist in a band with non-zero spectral winding, as argued in Ref.~\cite{zhang2020}.
In such a band, two states with opposite group velocities $v_g=\partial_{k_y}\Re E$ will have different imaginary energies, which determine their growth or decay rate during time evolution \cite{lee2019, longhi2022, kawabata2023}.
Therefore, at long timescales, the system develops a current, that is equal to the group velocity of the occupied state with the largest $\Im E$.
{In the 2D SSH model, the Zak phase protects edge states inside the spectral gap \cite{liu2017, obana2019}, as illustrated in Fig.~\ref{fig:edge_chirality}(e) where bulk vs.~edge states are shown with purple vs.~yellow color. Edge states form a band with} spectral winding when transition rates become non-reciprocal $c\neq0$ (see Fig.~\ref{fig:edge_chirality}(e)). This spectrum is an example of a topologically protected edge spectral winding, discussed in Ref.~\cite{ou2023}.  
Assuming that the quantum SSH model has one electron per unit cell, {the Fermi level goes through the bulk gap and intersects the edge state (Fig.~\ref{fig:edge_chirality}(e)).} The persistent current is calculated in the {occupied state with the largest $\Im E$} indicated with a dashed rectangle in Fig.~\ref{fig:edge_chirality}(e). Note that edge localization is calculated as the sum over squared magnitudes for all vector components lying on the edge $\sum_{j\in\textnormal{edge}}|\psi_j|^2$.

In the stochastic model, the diagonal terms $\mathcal{D}=\mathcal{A}-\mathcal{W}$ shift the edge band such that it overlaps with the bulk. As we showed above for Hermitian models, such shift leads to hybridization between edge and bulk states and a uniform steady state. However, as we will demonstrate in the last section of this manuscript, complex spectra in non-Hermitian models can decrease hybridization to allow edge states. The steady state in the studied 2D model, indicated with a dashed rectangle in Fig.~\ref{fig:edge_chirality}(f), is therefore a mixture of bulk distributed and edge localized probability.
The persistent current is calculated as a steady state probability flow along the network edge \cite{tang2021} and has a non-zero contribution only from the edge localized portion of the steady state (see Appendix \ref{app:stoch_2Dsurface} for details on the probability current).

Having identified the quantum and stochastic persistent currents, we track in Fig.~\ref{fig:edge_chirality}(g) how their values change with non-reciprocity $c$.
Again, the stochastic current grows dramatically with non-reciprocity, while the quantum current plateaus, just as in the 1D model. 
These trends can be understood by observing that non-reciprocity mainly affects the imaginary component of
the edge band, much more than the real component. As the quantum current depends on the momentum derivative of this real component, it is not strongly altered by non-reciprocity. Instead, the edge localized portion of the steady state grows with non-reciprocity \cite{tang2021}, which strongly affects the stochastic current. 

{We have found that stochastic systems depend far more radically on non-reciprocity compared to quantum systems, and both systems have different physical mechanisms for edge states. In the next section, we discuss how non-reciprocity leads to novel theoretical and physical consequences in stochastic systems, that underscore the value of our findings on non-reciprocity for a full understanding of stochastic systems.}

\section{Consequences of non-Hermiticity}

{Identifying the explicitly non-reciprocal conditions leads to the correct non-Hermitian classification, with new distinctions between different types of complex spectra and topological invariants \cite{kawabata2019}, many of which have no Hermitian counterpart. Topological classification becomes that of non-Hermitian systems, which is governed by 38 classes of non-Hermitian symmetry groups \cite{kawabata2019}, in contrast to 10 Hermitian symmetry groups \cite{schnyder2008,ryu2010}. Non-Hermitian topology is further classified by its spectral properties, i.e. point vs line gaps \cite{kawabata2019}, which are not distinguished in Hermitian spectra. }

{Using the correct classification leads to the appropriate topological invariant and prediction of relevant physical observables. For instance, point gap topology introduces a new non-Hermitian topological invariant, the spectral winding. Indeed, topological invariants in Hermitian systems are calculated using eigenvectors \cite{hasan2010, bansil2016}, since only eigenvectors are complex in Hermitian systems. In non-Hermitian systems, the complex eigenvalues allow for the calculation of topological invariants which have no counterpart in Hermitian systems \cite{gong2018, kawabata2019}. This is shown in our example of the 1D SSH model (Sec.~\ref{sec:NH_necessary}). In line gap topology, Hermitian topological invariants which are calculated using only right eigenvectors, are inadequate for non-Hermitian systems. Instead, a biorthogonal approach containing both left and right eigenvectors is necessary to provide a complete basis for non-Hermitian systems \cite{kunst2018}. An example of this is our 2d model (Sec.~\ref{sec:non-recipr_growth}), where the Zak phase must be calculated using such a non-Hermitian approach \cite{lieu2018,zheng2024}.}

{Besides key methodological differences in classification, the novel topological invariants lead to very different boundary sensitivity in Hermitian vs non-Hermitian systems. Boundary sensitivity only occurs in some non-Hermitian systems, those with point-gap spectra. This gives rise to the skin effect, where edge states scale with volume, in contrast to Hermitian edge states that scale with the boundary.} {In 1D systems, this boundary sensitivity occurs whenever the spectrum has a winding under periodic boundary conditions \cite{okuma2020}, and in both quantum and stochastic systems.} {In all these examples, we see that explicitly identifying the non-Hermitian nature of these topological phases and correct invariant is necessary for understanding the emergent physics.}

{While non-Hermitian matrices were used in prior studies \cite{dasbiswas2018, knebel2020}, these works mapped their matrices onto Hermitian matrices and used Hermitian topological classification and invariants. However, such an approach prevents connecting to concepts and results from non-Hermitian topology. For instance, it can miss the correct physical observables associated with non-Hermitian systems, such as the skin effect.} {
Although a Hermitian mapping can predict skin modes, it does so using semi-infinite boundary conditions \cite{okuma2020}. However, the semi-infinite boundary spectrum is larger than the spectrum of a fully open system, and hence, overestimates the number of skin modes. In quantum systems with line gaps, the non-Hermitian phase can be mapped onto its Hermitian counterpart
\cite{kawabata2019}. 
However, we showed above that this correspondence breaks down for  stochastic Hermitian systems.
}

{To illustrate the novelty that non-Hermiticity brings to symmetry classification and its interplay with invariants and physical observables, we detail these aspects for both the quantum and stochastic models. Our 1D model has time reversal symmetry and thus belongs to the AI symmetry class. Its spectrum has a point gap and so yields the $\mathbb{Z}$ classification in both quantum and stochastic systems \cite{kawabata2019, okuma2020}. An additional sublattice symmetry in the quantum system and a line gap give rise to the $\mathbb{Z}$ classification, which is responsible for zero-energy edge states \cite{yao2018, kawabata2019}. Our 2D model has inversion symmetry and hence falls outside the 38 symmetry classes that include only internal symmetries \cite{kawabata2019}. A system with inversion symmetry has a $\mathbb{Z}_2$ line-gap classification given by the Zak phase invariant \cite{liu2017, obana2019}, which has to be computed using biorthogonal eigenvectors in the non-Hermitian case \cite{lieu2018,zheng2024}. This classification applies to both quantum and stochastic systems. {A systematic comparison between the 1D and the 2D quantum and stochastic models is provided in Table~\ref{tab:models_sym_class} in the Appendix.}}

\section{Non-reciprocity restores the bulk-edge correspondence in stochastic systems}

\label{sec:edge_mechanism}

\begin{figure}
    \centering
    \includegraphics{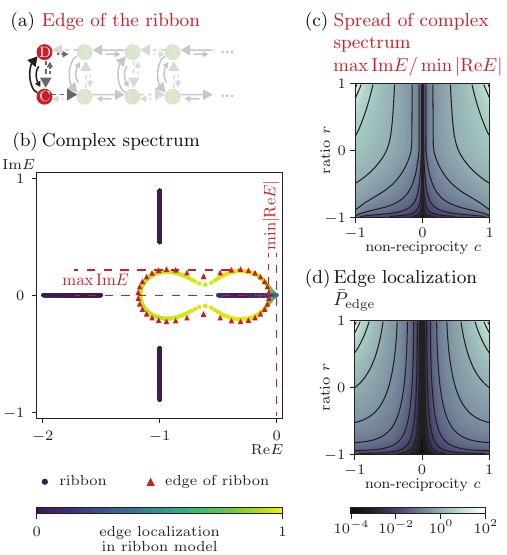}
    \caption{\textbf{In stochastic systems, spread of the complex spectrum reduces hybridization and induces edge localization, which is necessary for the edge current.}
    (a) To characterize the spread of the edge band, we introduce an auxiliary 1D model, that emulates the edge of the stochastic 2D SSH ribbon presented in Fig.~\ref{fig:edge_chirality}(d). 
    (b) To confirm that the two models are close, we plot the spectra of the 2D SSH model (purple vs.~yellow dots denoting bulk distributed vs.~edge localized states), and of the edge model (red triangles) at $r=0.5$ and $c=0.9$.
    We characterize the edge model spectrum by the largest imaginary component among all eigenvalues $\max\Im E$ and the smallest magnitude among all real components $\min|\Re E|$. We observe that (c) the overall complex spread of the edge model, expressed as $\max\Im E/\min|\Re E|$, is similar in shape to (d) total excess of probability on the edge of the 2D model $\bar{P}_\textnormal{edge}$. This similarity indicates that spread in the complex spectrum leads to edge localization and hence the steady state current.}
    \label{fig:quasi-edge}
\end{figure}

We showed that in the 1D SSH chain,
non-reciprocity localizes all eigenstates on one edge (Fig.~\ref{fig:SSH_spectrum_states}(f)).
{This behavior is due to the non-Hermitian skin effect and occurs in both quantum \cite{okuma2020} and stochastic \cite{sawada2024a, sawada2024b} models with
point gap topology, which has a unique boundary sensitivity. In contrast, line gap topology has been less well-studied, so in this section we explore a mechanism for how topological edge states emerge in such models.}
One clue comes from Fig.~\ref{fig:edge_chirality}(f), where we see that the bulk (purple) and edge (yellow) states become hybridized (green) close to the steady state, in contrast to the quantum system in Fig.~\ref{fig:edge_chirality}(e) where the bulk and edge states remain distinct.
At the same time this hybridization does not completely destroys localization of the steady state, as was the case in Hermitian systems. Hence, non-Hermiticity provides
a mechanism to minimize such hybridization by causing the spectrum to become complex and allowing eigenvalues to spread out.

To investigate this, we quantify the complex shape of the edge band and compare it to the steady state localization. We note that directly characterizing the shape of the edge band is not possible as it mixes with the bulk band in the vicinity of the steady state. Hence, we introduce a 1D edge model that emulates the edge of the 2D model  (see Fig.~\ref{fig:quasi-edge}(a)). It consists of states and transition rates that are closest to the edge of the 2D model, as well as the rates that go from the edge to the bulk ($\gamma_\textnormal{in}'$ and $\gamma_\textnormal{in}$), as described by the transition matrix {in momentum space}
\begin{equation}
    \mathcal{W}_\textnormal{1D edge}=\begin{pmatrix}
        -\gamma_\textnormal{in}-\gamma_\textnormal{in}'-\gamma_\textnormal{ex}' & \gamma_\textnormal{in}' + \gamma_\textnormal{ex} e^{ik} \\
        \gamma_\textnormal{in} + \gamma_\textnormal{ex}' e^{-i k} & -\gamma_\textnormal{in}-\gamma_\textnormal{in}'-\gamma_\textnormal{ex}
    \end{pmatrix}.
\end{equation}

The spectrum of the 1D edge model, shown in Fig.~\ref{fig:quasi-edge}(b), overlaps with the edge-localized band of the 2D model in the topological phase (see Appendix~\ref{app:1D_edge_model}).
We characterize the shape of this 1D edge spectrum by generalizing the ratio between imaginary and real components of the first non-zero eigenvalue, which is the coherence of stochastic oscillations \cite{barato2017}.
Here, for the numerator we use the largest imaginary component among all eigenvalues $\max\Im E$, which quantifies the spread in imaginary space. 
For the denominator, we use the smallest magnitude among real components $\min|\Re E|$, 
whose inverse $(\min|\Re E|)^{-1}$ is the longest lifetime among all the modes. Hence, we use the ratio $\max\Im E/\min|\Re E|$ as a measure of the spread of the complex spectrum weighted by its lifetime, which is plotted in Fig.~\ref{fig:quasi-edge}(c).

We calculate the edge-localization in the 2D model as the total excess of probability on the edge over the uniform bulk value $\bar{P}_\textnormal{edge}=\sum_i(\bar{p}_i-\bar{p}_\textnormal{bulk})/\bar{p}_\textnormal{bulk}$ \cite{tang2021} (see Appendix~\ref{app:stoch_2Dsurface}) and plot it in Fig.~\ref{fig:quasi-edge}(d).
We observe that the two quantities in Figs.~\ref{fig:quasi-edge}(c) and (d) have a similar shape as a function of model parameters. They both grow from zero in the reciprocal limit $c=0$ to their peak values in the fully non-reciprocal limit $c=\pm1$ and also grow with the ratio $r$.
This suggests that the spread in the complex spectrum leads to edge localization by reducing hybridization between the edge and bulk eigenstates. Hence non-reciprocity, which allows edge states to be distinct from the bulk states in the complex plane, restores the bulk-edge correspondence even in the absence of the skin effect. 

{To understand if this mechanism is more general, we explore these metrics in another class of systems with a different symmetry group, using a 2D non-Hermitian model on the kagome lattice. We find that its edge probability as a function of model parameters is also similar to the complex spread of its edge spectrum weighted by its lifetime (details in Appendix~\ref{app:kagome}). This result is consistent with our findings for square lattice model, revealing the same mechanism for edge states at play in both systems governed by different symmetries.}

\section{Discussion}

Our results demonstrate the clear necessity and importance of non-reciprocity in stochastic topological systems. We showed two ways in which non-reciprocity mitigates the effects of hybridization towards robust edge responses in stochastic systems. We emphasize, that non-reciprocity condition is stronger than that of broken detailed balance.After all, detailed balance applies in 1d systems, including when the skin effect is present. In higher dimensions, detailed balance can be broken, such as in our 2D example where the steady state is a non-equilibrium current. Hence, our work covers examples both with and without detailed balance, showing that non-reciprocity is a stricter requirement to obtain a topological response, as it holds for all dimensions and types of edge states.

The necessity of non-reciprocity for stochastic topology could shed light on the observed behavior of real systems. Biological systems are often found to operate in a strongly non-equilibrium regime \cite{fang2019a}, despite the high energy consumption required to maintain these fluxes. Indeed, the observed energy consumption is often much higher than that estimated using thermodynamics calculations, on both microscopic \cite{hopfield1974,samoilov2005} and macroscopic levels \cite{cao2015,lan2012}. Our results suggest an intriguing reason for this strong non-equilibrium: non-reciprocity and its associated energy consumption is necessary for generically robust function. Towards building motile synthetic systems, our work also demonstrates the need for non-reciprocity in stabilizing the emergent behavior \cite{tang2021}. These conditions for specific and targeted responses, provide design principles for synthetic biological systems with robust behavior \cite{schwille2018, synthbio2017}, such as biochemical oscillators \cite{novak2008, li2018} or ring attractors \cite{churchland2012, khona2022}.

More broadly, we expect that the unique features of stochastic systems will lead to other modifications in the manifestations of topology, at times expanding and other times reducing the classification.
For example, stochastic matrices are restricted to symmetry classes that include time reversal symmetry $\mathcal{T}_+\mathcal{K}$, whose unitary part equals to identity $\mathcal{T}_+=\mathbbm{1}$ and hence squares to one $\mathcal{T}_+\mathcal{T}_+^*=\mathbbm{1}$, and $\mathcal{K}$ denotes the complex conjugation. Indeed, with real transition rates, any stochastic matrix $\mathcal{W}(k)$ fulfills $\mathcal{T}_+\mathcal{W}^*(k)\mathcal{T}_+^{-1}=\mathcal{W}(-k)$ with $\mathcal{T}_+=\mathbbm{1}$.
This presence of time-reversal symmetry reduces the number of relevant symmetry classes to only three out of eight real Altland-Zirnbauer symmetry classes, before further distinctions due to non-Hermiticity~\cite{kawabata2019}.
As there can be specific interactions between each symmetry group and the constraints of stochastic systems, this deserves to be explored individually for each class.

{The necessity of a non-Hermitian description shown in our work raises further questions about how various non-Hermitian phenomena affect the behavior of stochastic networks. We showed that 1D systems with point gaps produce edge states from the boundary sensitivity of the skin effect. Point gap topology in higher dimensions shows not only the skin effect \cite{hu2023, zhang2022, jiang2023} but also other responses, such as exceptional or infernal points \cite{denner2023}. Future study is needed to determine their role for edge states in stochastic systems. Furthermore, exceptional points, that are unique to non-Hermitian systems \cite{kawabata2019a, ding2022, heiss2012}, were shown to protect edge-localized states in quantum and hydrodynamic materials \cite{sone2020}. It would be interesting to understand the interplay of these topological features with the rich physics of stochastic systems. Our work clarifies general principles towards sustaining robust function in systems with a flux constraint, such as stochastic systems, or those governed by a Laplacian matrix, which can describe consensus or synchronization behavior \cite{newman2018networks}.}

\begin{acknowledgments}
\emph{Acknowledgments.---} 
We thank Jaime Agudo-Canalejo for helpful discussions and Abhijeet Melkani for comments on the manuscript.
A.N. and E.T. acknowledge support from the NSF Center for Theoretical Biological Physics (PHY-2019745) and E.T. acknowledges support from NSF CAREER Award (DMR-2238667).
\end{acknowledgments}

\appendix

\section{Su-Schriefer-Heeger model}

In this appendix, we provide more details about the 1D Su-Schriefer-Heeger (SSH) model and its topological classification. Section \ref{app:SSH_topology} deals with the Hermitian model, while section \ref{app:SSH_NH} with the non-Hermitian model. We derive the zero eigenstates of the non-Hermitian SSH model in section \ref{app:IPR}.

\subsection{Hermitian SSH model}
\label{app:SSH_topology}

Hermitian SSH model is shown in Fig.~\ref{fig:SSH_spectrum_states}(a) in the main text. Due to translational symmetry, under periodic boundary conditions this model is described by a matrix defined over reciprocal space. The quantum Hamiltonian as a function of momentum $k\in[-\pi,\pi]$ is given by 
\begin{equation}
    \mathcal{A} = 
    \begin{pmatrix}
        0 & \gamma_\textnormal{in} + \gamma_\textnormal{ex} e^{ik} \\
        \gamma_\textnormal{in} + \gamma_\textnormal{ex} e^{-ik} & 0
    \end{pmatrix}.
\end{equation}
The corresponding stochastic transition matrix includes additional diagonal terms that account for probability conservation:
\begin{equation}
    \mathcal{W}=
    \begin{pmatrix}
        -\gamma_\textnormal{in}-\gamma_\textnormal{ex} & \gamma_\textnormal{in} + \gamma_\textnormal{ex} e^{ik} \\
        \gamma_\textnormal{in} + \gamma_\textnormal{ex} e^{-ik} & -\gamma_\textnormal{in} -\gamma_\textnormal{ex}
    \end{pmatrix}.
\end{equation}
With periodic boundary conditions, the quantum and stochastic matrices differ merely by a constant shift and therefore have the same topological classification. Namely, they exist in two different phases, determined by the Zak phase topological invariant \cite{berry1983, zak1989} protected by the inversion symmetry $\sigma_x\mathcal{M}(k)\sigma_x=\mathcal{M}(-k)$ for $\mathcal{M}=\mathcal{A},\mathcal{W}$. The Zak phase takes the form  
\begin{equation}
    \Phi=\int_{-\pi}^{\pi} dk \bra{\psi_1(k)}\partial_k\ket{\psi_1(k)}\in\{0, \pi\},
\end{equation}
where $\ket{\psi_1(k)}$ is the eigenvector of both quantum $\mathcal{A}(k)$ and stochastic $\mathcal{W}(k)$ matrices, corresponding to the lower eigenvalue. The Zak phase is non-zero when $\gamma_\textnormal{ex}>\gamma_\textnormal{in}$.

In the main text we showed that under open boundary conditions, the quantum and stochastic models have different behavior: the quantum model has zero-energy eigenstates localized on the edge, while the stochastic steady state is uniformly distributed over the bulk. In the main text, this result was proven generally for any Hermitian model. Here we provide an additional model-specific explanation for the absence of edge-localized stochastic modes. We know that the stochastic SSH model is different from the quantum SSH model by the diagonal terms $\mathcal{W}=\mathcal{A}-\mathcal{D}$. We notice that under open boundary conditions, the diagonal terms are decomposed into a constant diagonal and an edge-localized matrix $\mathcal{D}=(\gamma_\textnormal{in}+\gamma_\textnormal{ex})I-\mathcal{P}$. Here $I$ is the identity matrix and
\begin{equation}
\mathcal{P}_{ij, \alpha\beta}=(\delta_{i1}\delta_{j1}\delta_{\alpha 1}\delta_{\beta1}+\delta_{iN}\delta_{jN}\delta_{\alpha 2}\delta_{\beta2})\gamma_\textnormal{ex}, 
\end{equation}
with $i,j=1, ..N$ being unit cell indices and $\alpha, \beta=1, 2$ indicating a state within the unit cell. The edge-localized term $\mathcal{P}$  breaks the sublattice symmetry $\sigma_z\mathcal{A}(k)\sigma_z=-\mathcal{A}(-k)$, which protects the zero-energy states in the quantum model \cite{asboth2016}. Therefore, edge-localized states do not exist in the gap of the stochastic model due to the symmetry-breaking edge terms.

\subsection{Non-Hermitian SSH model}
\label{app:SSH_NH}

The non-Hermitian SSH model is illustrated in Fig.~\ref{fig:edge_chirality}(a) in the main text and generalizes the Hermitian SSH model by allowing non-reciprocal transition rates. The quantum model under periodic boundary conditions is given by 
\begin{equation}
    \mathcal{A} = 
    \begin{pmatrix}
        0 & \gamma_\textnormal{in}' + \gamma_\textnormal{ex} e^{ik} \\
        \gamma_\textnormal{in} + \gamma_\textnormal{ex}' e^{-ik} & 0
    \end{pmatrix}.
\end{equation}
where $\gamma_\textnormal{in}'$ and $\gamma_\textnormal{ex}'$ are backwards rates corresponding to the forward rates $\gamma_\textnormal{in}$ and $\gamma_\textnormal{ex}$, respectively. The stochastic model is given by
\begin{equation}
    \mathcal{W}=
    \begin{pmatrix}
        -\gamma_\textnormal{in}-\gamma_\textnormal{ex}' & \gamma_\textnormal{in}' + \gamma_\textnormal{ex} e^{ik} \\
        \gamma_\textnormal{in} + \gamma_\textnormal{ex}' e^{-ik} & -\gamma_\textnormal{in}' -\gamma_\textnormal{ex}
    \end{pmatrix},
\end{equation}
To reduce the parameter space, the four rates are parameterized using the overall scaling factor $\gamma_\textnormal{tot}$, ratio $r$ and non-reciprocity $c$: 
\begin{align}
& \gamma_\textnormal{ex}=\gamma_\textnormal{tot}(1+c)(1+r)/4, \nonumber \\ &\gamma_\textnormal{ex}'=\gamma_\textnormal{tot}(1-c)(1+r)/4, \nonumber \\ &\gamma_\textnormal{in}=\gamma_\textnormal{tot}(1+c)(1-r)/4, \nonumber \\
&\gamma_\textnormal{in}'=\gamma_\textnormal{tot}(1-c)(1-r)/4.
\label{eq:rate_parametrization}
\end{align}

Both the quantum and stochastic models {have the time-reversal symmetry and therefore belong to the AI class of non-Hermitian symmetries \cite{kawabata2019}, as specified in Table~\ref{tab:models_sym_class}. In this symmetry class, point gap topology of these 1D models is characterized by } a non-Hermitian topological invariant -- the spectral winding number
\begin{equation}
    \nu(E_0)=\frac{1}{2\pi i}\int_{-\pi}^\pi\partial_k \log(\det(\mathcal{M}(k)-E_0)) dk,
\end{equation}
with $\mathcal{M}=\mathcal{A}$ in the quantum case and $\mathcal{M}=\mathcal{W}$ in the stochastic case. $E_0$ is a reference energy around which the spectrum winds. For all parameters away from $r=\pm1$ or $c=0$, both quantum and stochastic bands produce a non-zero winding.
This means that both systems exhibit the skin effect \cite{okuma2020, sawada2024a, sawada2024b} and have their states (including the zero-energy states) localized on one edge.

{Additional sublattice symmetry in the quantum system leads to a line gap topology with $\mathbb{Z}$ classification given by a non-Bloch winding number \cite{yao2018}. The studied 1D model distinguishes between two line-gap phases, with winding number equal to zero and one. When the ratio $r>0$, the winding number equals to one, and the system has zero-energy edge states, illustrated in Fig.~\ref{fig:SSH_spectrum_states}(e).  We derive these zero-energy eigenstates in the next subsection.} 

{\renewcommand{\arraystretch}{2}
\begin{table*}
    \centering
    \begin{tabular}{r|r|l|l}
    \hline\hline
         && quantum & stochastic  \\
         \hline
         1D & symmetries & \makecell[lt]{time-reversal  \\
         $\mathcal{T}_+\mathcal{A}^*(k)\mathcal{T}_+^{-1}=\mathcal{A}(-k)$
         \smallskip
         \\  sublattice \\
         $\mathcal{S}\mathcal{A}(k)\mathcal{S}^{-1}=-\mathcal{A}(k)$} & \makecell[lt]{time-reversal \\
         $\mathcal{T}_+\mathcal{W}^*(k)\mathcal{T}_+^{-1}=\mathcal{W}(-k)$
         } \\
         & symmetry class & \makecell[lt]{AI with $\mathcal{S}^+$ \cite{kawabata2019} \\($\mathcal{T}_+$ and $\mathcal{S}$ commute)} & AI \cite{kawabata2019} \\
         & classification & 
         $\mathbb{Z}\oplus\mathbb{Z}$ \cite{kawabata2019}
         & $\mathbb{Z}$ \cite{kawabata2019} \\
         \hline
         2D & symmetries & \makecell[lt]{time-reversal \\
         $\mathcal{T}_+\mathcal{A}^*(k)\mathcal{T}_+^{-1}=\mathcal{A}(-k)$
         \smallskip
         \\
         inversion \\
         $\mathcal{I}\mathcal{A}(k)\mathcal{I}^{-1}=\mathcal{A}(-k)$
         \smallskip \\
         sublattice \\
         $\mathcal{S}\mathcal{A}(k)\mathcal{S}^{-1}=-\mathcal{A}(k)$} & \makecell[lt]{time-reversal \\
         $\mathcal{T}_+\mathcal{W}^*(k)\mathcal{T}_+^{-1}=\mathcal{W}(-k)$
         \smallskip
         \\
         inversion \\
         $\mathcal{I}\mathcal{W}(k)\mathcal{I}^{-1}=\mathcal{W}(-k)$}\\
         & symmetry class & Zak phase \cite{liu2017} & Zak phase \cite{liu2017} \\
         & classification & $\mathbb{Z}_2$ \cite{liu2017} & $\mathbb{Z}_2$ \cite{liu2017} \\
         \hline\hline
    \end{tabular}
    \caption{Symmetry class and topological classification of quantum and stochastic 1D and 2D models. For the matrix representations we use, the operators can take the form $\mathcal{T}_+=\mathbbm{1}$ and $\mathcal{S}=\sigma_z$ in the 1d models, and $\mathcal{T}_+=\mathbbm{1}$, $\mathcal{S}=\mathbbm{1}\otimes\sigma_z$ and $\mathcal{I}=\sigma_x\otimes\mathbbm{1}$ in the 2d models.} 
    \label{tab:models_sym_class}
\end{table*}}

\subsection{Zero eigenstates and inverse participation ratio}
\label{app:IPR}

We proceed to derive the zero-energy eigenstates of the quantum and stochastic SSH models and compute their inverse participation ratio. We follow the formalism presented for the quantum SSH model in Ref.~\cite{jiang2020}.

\subsubsection{Quantum zero-energy eigenstates}

First, consider the quantum model and reproduce the derivation from Ref.~\cite{jiang2020}. We represent the system in real space and write the wavefunctions in the basis of vectors $\ket{n}\otimes\ket{\xi}$, with $\ket{n}$ encoding a state in the $n$-th unit cell, and $\ket{\xi}=(\xi_1,\xi_2)^T$ specifying the state distribution over sublattice $1$ and $2$. The quantum SSH Hamiltonian in real space is given by
\begin{align}
    \hat{\mathcal{A}}=&\sum_n\begin{pmatrix}
        0 & \gamma_\textnormal{in}' \\
        \gamma_\textnormal{in}  & 0
    \end{pmatrix}\ket{n}\bra{n}\nonumber\\
    &+\begin{pmatrix}
        0 &  \gamma_\textnormal{ex} \\
        0 & 0
    \end{pmatrix} \ket{n}\bra{n-1}
    + \begin{pmatrix}
        0 & 0 \\
         \gamma_\textnormal{ex}' & 0
    \end{pmatrix}\ket{n}\bra{n+1}
\end{align}
To focus on a single edge, we assume that the chain is semi-infinite with only left edge. We search for an edge-localized eigenstate using the ansatz
\begin{equation}
    \ket{\psi_l}=\frac{1}{\mathcal{N}}\sum_{n=1}^N\beta^{n-1}\ket{n}\otimes\ket{\xi},
    \label{eq:left_evec}
\end{equation}
with the size of the system $N\to\infty$.
This state is a zero-energy eigenstate of the Hamiltonian if $\mathcal{A}\ket{\psi_l}=0$. We also search for a corresponding right eigenstate $\mathcal{A}^\dagger\ket{\phi_l}=0$, using the ansztz
\begin{equation}
    \ket{\phi_l}=\frac{1}{\mathcal{N}}\sum_{n=1}^N\beta'^{n-1}\ket{n}\otimes\ket{\xi}.
\end{equation}
We require that the two states form a biorthogonal pair $\bra{\phi_l}\ket{\psi_l}=1$.

Substituting the anzatz \eqref{eq:left_evec} to the eigenstate equation, we get the following equation in the bulk of the chain
\begin{equation}
    \begin{pmatrix}
        0 & \gamma_\textnormal{in}' + \gamma_\textnormal{ex} \beta^{-1} \\
        \gamma_\textnormal{in} + \gamma_\textnormal{ex}'\beta & 0
    \end{pmatrix} \ket{\xi}=0.
\end{equation}
On the left edge the same vector $\ket{\xi}$ must satisfy the boundary condition
\begin{equation}
    \begin{pmatrix}
        0 & \gamma_\textnormal{in}' \\
        \gamma_\textnormal{in} + \gamma_\textnormal{ex}'\beta & 0
    \end{pmatrix} \ket{\xi}=0.
\end{equation}
These equations are solved by $\beta=-\gamma_\textnormal{in}/\gamma_\textnormal{ex}'$ and $\ket{\xi}=(1, 0)^T$. Similarly, from the right eigenstate equation we get $\beta'=-\gamma_\textnormal{in}'/\gamma_\textnormal{ex}$. Biorthogonality of the two states specifies the normalization factor to be $\mathcal{N}=\sqrt{(1-(\beta'\beta)^N)/(1-\beta'\beta)}$. For the zero-energy state to exist in a semi-infinite chain, the normalization factor must be finite in the limit of $N\to\infty$, which is true if $|\beta\beta'|<1$. The zero-energy state forms a decaying solution when $|\beta|<1$. Expressing the rates in terms of the ratio $r$ and non-reciprocity $c$, as in Eq.~\eqref{eq:rate_parametrization}, the left-localized zero-energy state exists for $r>0$ (e.g. when external rates are larger than internal) and $r>c$.

Following the same procedure, we derive the  right-localized zero-energy state in a semi-infinite chain with the right edge. It is given by
\begin{equation}
    \ket{\psi_r}=\frac{1}{\mathcal{N}}\sum_{n=0}^{N-1}\beta'^{n}\ket{N-n}\otimes\ket{\xi'},
    \label{eq:right_evec}
\end{equation}
with $\ket{\xi'}=(0, 1)^T$. This state exists and has a finite normalization factor $\mathcal{N}$ when $r>0$ and $r>-c$. In Fig.~\ref{fig:NHSSH_right_left} we illustrate parameter regions, in which a semi-infinite model has a left- or right-localized eigenstate with green and purple filling, respectively.

\begin{figure}
    \centering
    \includegraphics{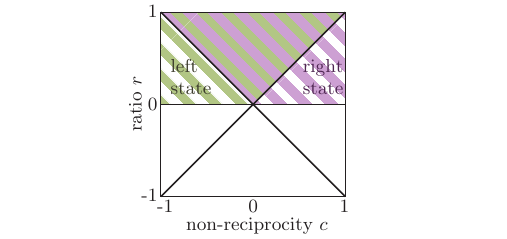}
    \caption{Green vs. purple filling indicates parameter values $c$ and $r$, for which left- vs.~right-localized zero-energy state is present in a semi-infinite quantum SSH chain.}
    \label{fig:NHSSH_right_left}
\end{figure}

In a finite chain with both left and right edges, the eigenstates close to the zero energy are formed by a linear combination of right- and left-localized eigenstates \cite{jiang2020}. The contribution of the right (left) eigenstate decays exponentially with the chain length $N$ for $c<0$ ($c>0$). Therefore, in the following we assume that when $c<0$, $r>0$ the model has a left-localized zero-energy state \eqref{eq:left_evec}, and when $c>0$, $r>0$ a right-localized state \eqref{eq:right_evec}. For $r<0$ no normalized zero-energy state can be found.

Next, to quantify the degree of localization of these states we compute their inverse participation ratio as defined in the main text
\begin{align}
    \mathrm{IPR}^\mathrm{q}_l&=\frac{\sum_{j=1}^N|\psi_{l,j}|^4}{\bra{\psi_l}\ket{\psi_l}^2} \nonumber \\
    &=\sum_{n=1}^{\infty}\frac{|\beta^{n-1}|^4}{\left(\sum_{n=1}^{\infty} |\beta^{n-1}|^2\right)^2}\nonumber\\
    &=\frac{\left(1-(\gamma_\textnormal{in}/\gamma_\textnormal{ex}')^2\right)^2}{1-(\gamma_\textnormal{in}/\gamma_\textnormal{ex}')^4} \nonumber \\
    &=\frac{(1-c)^2(1+r)^2-(1+c)^2(1-r)^2}{(1-c)^2(1+r)^2+(1+c)^2(1-r)^2}.
\end{align}
\begin{equation}
    \mathrm{IPR}^\mathrm{q}_r=\dots=\frac{(1+c)^2(1+r)^2-(1-c)^2(1-r)^2}{(1+c)^2(1+r)^2+(1-c)^2(1-r)^2}.
\end{equation}
We further define a directional inverse participation ratio (dIPR) whose sign indicates whether state is localized on the left or on the right
\begin{equation}
    \mathrm{dIPR}=\begin{cases}
        -\mathrm{IPR}_l, & c<0 \\
        \mathrm{IPR}_r, & c>0
    \end{cases}
\end{equation}
In Figure~\ref{fig:NHSSH_dIPR}(a) we show dIPR for the full range of parameters $c$ and $r$.

\begin{figure}
    \centering
    \includegraphics{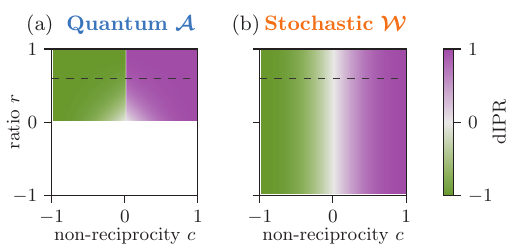}
    \caption{Directional inverse participation ratio in a long non-Hermitian SSH chain for a range of ratio $r$ and non-reciprocity $c$. (a) dIPR computed for the zero-energy quantum state. (b) dIPR computed for the stochastic steady state. Dashed line indicates the line, along which the inverse participation ratio is plotted in the main text in Fig.~\ref{fig:edge_chirality}(b).}
    \label{fig:NHSSH_dIPR}
\end{figure}

\subsubsection{Stochastic steady state}

We proceed to derive the steady state in the stochastic model. In real space, the transition matrix takes the form
\begin{align}
    \hat{\mathcal{W}}=&\sum_n\begin{pmatrix}
        -\gamma_\textnormal{in}-\gamma_\textnormal{ex}' & \gamma_\textnormal{in}' \\
        \gamma_\textnormal{in}  & -\gamma_\textnormal{in}'-\gamma_\textnormal{ex}
    \end{pmatrix}\ket{n}\bra{n}\nonumber\\
    &+\begin{pmatrix}
        0 &  \gamma_\textnormal{ex} \\
        0 & 0
    \end{pmatrix} \ket{n}\bra{n-1}
    + \begin{pmatrix}
        0 & 0 \\
         \gamma_\textnormal{ex}' & 0
    \end{pmatrix}\ket{n}\bra{n+1}.
\end{align}
Similar to the quantum case, we search for a left-localized zero eigenstate of the transition matrix $\hat{\mathcal{W}}\ket{p_l}=0$ in the form
\begin{equation}
    \ket{p_l}=\frac{1}{\mathcal{N}}\sum_{n=1}^N\delta^{n-1}\ket{n}\otimes\ket{\chi}
\end{equation}
In the bulk this state must fulfill
\begin{equation}
    \begin{pmatrix}
        -\gamma_\textnormal{in}-\gamma_\textnormal{ex}' & \gamma_\textnormal{in}'  + \gamma_\textnormal{ex}\delta^{-1}\\
        \gamma_\textnormal{in} + \gamma_\textnormal{ex}'\delta & -\gamma_\textnormal{in}'-\gamma_\textnormal{ex}
    \end{pmatrix} \ket{\chi}=0,
\end{equation}
while the left boundary condition reads
\begin{equation}
    \begin{pmatrix}
        -\gamma_\textnormal{in} & \gamma_\textnormal{in}' \\
        \gamma_\textnormal{in} + \gamma_\textnormal{ex}'\delta & -\gamma_\textnormal{in}'-\gamma_\textnormal{ex}
    \end{pmatrix} \ket{\chi}=0.
\end{equation}
The state that satisfies this condition is given by $\delta=\gamma_\textnormal{in}\gamma_\textnormal{ex}/\gamma_\textnormal{in}'\gamma_\textnormal{ex}'$ and $\ket{\chi}=(\gamma_\textnormal{in}', \gamma_\textnormal{in})^T$. This state decays from the left edge if $|\gamma_\textnormal{in}\gamma_\textnormal{ex}|<|\gamma_\textnormal{in}'\gamma_\textnormal{ex}'|$, which in terms of paramters $c$ and $r$ means $c<0$. Computing the corresponding right eigenvector $\mathcal{W}^\dagger\ket{p_l'}=0$, we find that the binormalization factor $\mathcal{N}$ is finite for all $r\neq\pm1$ and $c\neq0$. 

Similarly, the right-localized eigenstate can be found using the ansatz 
\begin{equation}
    \ket{p_r}=\frac{1}{\mathcal{N}}\sum_{n=0}^{N-1}\delta'^{n}\ket{N-n}\otimes\ket{\chi'}.
\end{equation}
Substituting it to the eigenstate equation we get $\delta'=\delta^{-1}$ and $\ket{\chi'}=\ket{\chi}$, which decays from the right edge when $c>0$.

Finally, we compute the IPR for these eigenstates. When $c<0$, IPR of the left-localized eigenstate is
\begin{align}
    \mathrm{IPR}_l^\mathrm{st}&=\frac{\sum_{j=1}^N|p_{l,j}|^4}{\bra{p_l}\ket{p_l}^2} \nonumber \\&=\sum_{n=1}^{\infty}\frac{|\delta^{n-1}|^4(|\chi_1|^4+|\chi_2|^4)}{\left(\sum_{n=1}^{\infty} |\delta^{n-1}|^2(|\chi_1|^2+|\chi_2|^2)\right)^2}\nonumber\\
    &=\frac{((1-c)^4+(1+c)^4)\left(1-(1+c)^4/(1-c)^4\right)^2}{((1-c)^2+(1+c)^2)^2\left(1-(1+c)^8/(1-c)^8\right)} \nonumber \\
    &=\frac{(1-c)^2-(1+c)^2}{(1-c)^2+(1+c)^2}=-\frac{2c}{1+c^2}.
\end{align}
Similarly, when $c>0$, the IPR of the right localized eigenstate is
\begin{equation}
    \mathrm{IPR}_r^\mathrm{st}=\frac{2c}{1+c^2}.
\end{equation}
We plot the corresponding dIPR for a range of parameters $c$ and $r$ in Fig.~\ref{fig:NHSSH_dIPR}(b). In the main text, the absolute value of dIPR for the quantum and stochastic models is shown along a one-dimensional cut $r=0.6$, indicated with a dashed line in Fig.~\ref{fig:NHSSH_dIPR}(a) and (b).

\section{2D SSH model}
\label{app:SSH2D}

In this Appendix, we provide more details about the 2D non-Hermitian Su-Schrieffer-Heeger model. In Sec.~\ref{app:2DSSH_matr} we specify the quantum Hamiltonian and the stochastic transition matrix. We discuss their topological classification in Sec.~\ref{app:2DSSH_topo}. In Sec.~\ref{app:stoch_2Dsurface}, we introduce the stochastic steady state probability and the edge current in a ribbon geometry. Finally, in Sec.~\ref{app:1D_edge_model}, we describe the 1D edge model that emulates the edge of the 2D model.

\subsection{Transition matrices}
\label{app:2DSSH_matr}

The 2D SSH network \cite{tang2021} is shown in Fig.~\ref{fig:edge_chirality}(d) and consists of four states per unit cell. The states are connected by internal transition rates $\gamma_\textnormal{in}$, which form a unidirectional cycle. Each forward rate is complemented with the corresponding backward rate $\gamma_\textnormal{in}'$. The unit cells are connected by external transition rates $\gamma_\textnormal{ex}$, that form a cycle in the opposite direction with respect to the internal cycle. The corresponding backward rates are $\gamma_\textnormal{ex}'$. In momentum space, the quantum 2D SSH model is described by the following Hamiltonian
    \begin{equation}
        \begin{aligned}
\mathcal{A}=&
\left(\begin{matrix}
            0 & \gamma_\textnormal{in}+\gamma'_\textnormal{ex}e^{-ik_y} \\
            \gamma'_\textnormal{in} + \gamma_\textnormal{ex}e^{ik_y} & 0 \\
            0 & \gamma'_\textnormal{in} + \gamma_\textnormal{ex}e^{ik_x} \\
            \gamma_\textnormal{in}+\gamma'_\textnormal{ex}e^{ik_x} & 0
         \end{matrix} \right.
         \\
&\qquad\qquad
\left.\begin{matrix}
    0 & \gamma'_\textnormal{in} + \gamma_\textnormal{ex}e^{-ik_x} \\
    \gamma_\textnormal{in}+\gamma'_\textnormal{ex}e^{-ik_x} & 0  \\
    0 & \gamma_\textnormal{in}+\gamma'_\textnormal{ex}e^{ik_y} \\
    \gamma'_\textnormal{in} + \gamma_\textnormal{ex}e^{-ik_y} & 0
\end{matrix}
\right).
\end{aligned}
\label{eq:SSH_2D}
\end{equation}
The stochastic transition matrix on the same network is given by
\begin{equation}
    \mathcal{W} = \mathcal{A} - \gamma_\textnormal{tot}I,
\end{equation}
where $I$ is the identity matrix, and $\gamma_\textnormal{tot}=\gamma_\textnormal{in}+\gamma_\textnormal{in}'+\gamma_\textnormal{ex}+\gamma_\textnormal{ex}'$. The second term ensures probability conservation in the stochastic model. In the following we assume the transition rates to be parameterized by non-reciprocity $c$ and ratio $r$, as in Eq.~\eqref{eq:rate_parametrization}.

\begin{figure*}
    \centering
    \includegraphics{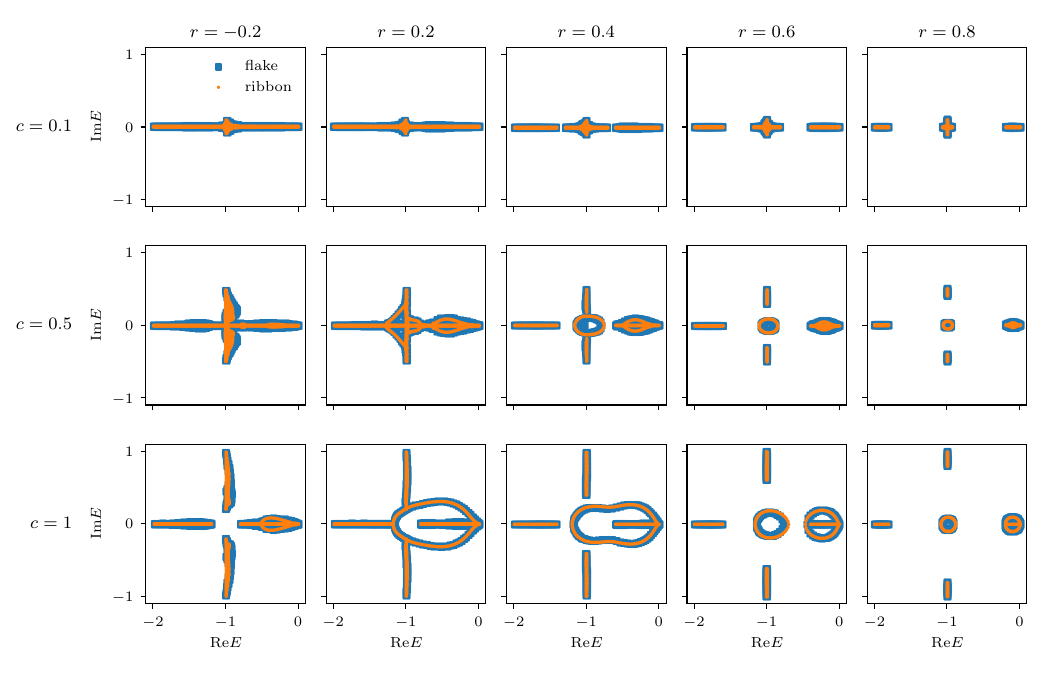}
    \caption{Comparison of the 2D SSH spectrum in the flake geometry with all open boundaries (blue squares) and in the ribbon geometry with periodic boundary condition along the $y$-axis (orange dots) for a number of model parameters $c$ and $r$. The spectra are similar, indicating that the ribbon geometry is a good approximation for a fully open system.}
    \label{fig:ribbon_vs_flake}
\end{figure*}

In the main text, we study how this model behaves on the edge. We stated that a model with open boundary conditions in both $x$ and $y$ directions can be approximated with a model in ribbon geometry with periodic boundary conditions in the $y$ direction. To justify this approximation, we plot in Fig.~\ref{fig:ribbon_vs_flake} the spectra of the fully open model (with blue squares) and the ribbon model (with orange dots) for a range of parameters $c$ and $r$. All models have 10 unit cells in the directions with open boundary conditions. We observe that the spectra of these models are close to each other, justifying our approximation.

\subsection{Topological classification}
\label{app:2DSSH_topo}

In this section, we discuss topological invariants in the 2D SSH model. 
We start from the Hermitian case, when the rates are reciprocal $c=0$, and follow Ref.~\cite{liu2017, obana2019} to characterize the topology of our model with a vectored Zak phase $(\Phi_x, \Phi_y)$ {protected by inversion symmetry}
\begin{align}
    \Phi_x&=\int_{-\pi}^{\pi} dk_x \bra{\psi_1(\bm{k})}\partial_{k_x} \ket{\psi_1(\bm{k})}\in\{0, \pi\}, \\
    \Phi_y&=\int_{-\pi}^{\pi} dk_y \bra{\psi_1(\bm{k})}\partial_{k_y} \ket{\psi_1(\bm{k})}\in\{0, \pi\},
\end{align}
where $\ket{\psi_1(\bm{k})}$ is the lowest eigenvector of both the quantum and the stochastic matrix. {Topological classification of this model is summarized in the lower row of Table~\ref{tab:models_sym_class}.}
For external rates being larger than internal rates $r>0$, the model is in a topological phase with $(\Phi_x, \Phi_y)=(\pi, \pi)$. 
When moving away from the  Hermitian limit $c\neq0$, the band structure becomes complex, as illustrated in Fig.~\ref{fig:edge_chirality}(e) of the main text for parameters $c=0.9$, $r=0.5$. If the absolute value of ratio remains larger than the critical value $|r|>r^*(c)$ \cite{tang2021}, the gap between the lower and the middle bulk bands remains open, and the system stays in the same topological phase as the Hermitian model. {To characterize this non-Hermitian phase, we can compute the Zak phase invariant using a biorthogonal basis \cite{lieu2018}.} 

We further discuss how this topological phase manifests on the edge of the quantum 2D SSH model. The Zak phase causes edge-localized states inside of the bulk gap. In the Hermitian limit, stability of these edge states was analyzed in Ref.~\cite{liu2017}, which stated that the edge states remain inside of the gap for perturbations with magnitude smaller than the gap size.
Non-reciprocal rates cause the edge states to have a spectral winding, which is demonstrated for the left-localized band in Fig.~\ref{fig:boundary_winding}, with momentum $k_y$ indicated by color to help visualizing the spectral winding. The winding number around the zero energy amounts to $\nu(E=0)=-1$. The right-localized bands are degenerate with the left-localized bands and have the opposite winding, as shown in the inset of Fig.~\ref{fig:boundary_winding}. As discussed in the main text, the winding of the edge bands leads to persistent currents on the edge of the quantum model. We plot this current for the full range of parameters $r$ and $c$, at which the model is in the insulating phase (Fig.~\ref{fig:current_cr}(a)). The critical ratio $\pm r^*(c)$, at which the model transitions to a semimetal is shown with a gray line.

\begin{figure}
    \centering
    \includegraphics{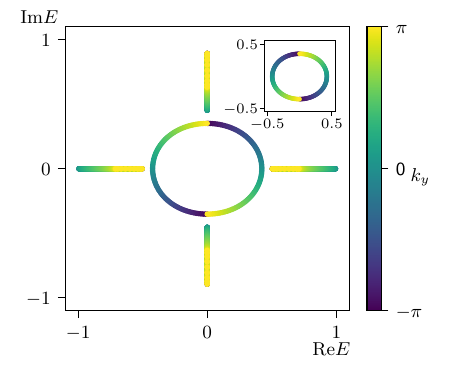}
    \caption{Spectrum of the quantum 2D non-Hermitian SSH ribbon with color denoting momentum $k_y$ of each eigenenergy. Momentum resolution helps to clearly see the winding of the edge bands in the middle. The main figure shows the left-localized band, while the right-localized band is shown in the inset and winds in the opposite direction. The plot is made at $r=0.9$ and $c=0.5$.}
    \label{fig:boundary_winding} 
\end{figure}

\subsection{Stochastic edge observables}
\label{app:stoch_2Dsurface}

In this section, we summarize the edge quantities computed in the stochastic 2D SSH model: the steady state probability and the edge-localized current. These quantities were derived in Ref.~\cite{tang2021} in Appendix D using the transfer matrix formalism, while here we reproduce the results.

The transfer matrix that acts on the probability vector in the $n$-th unit cell $[p_D^n, p_C^n, p_A^n, p_B^n]^T$ is given by
\begin{align}
    M&=\begin{pmatrix}
        U_1 & U_2 \\
        U_4U_1 & U_4U_2+U_3
    \end{pmatrix}, \\
    U_1&=\begin{pmatrix}
        -\gamma_\textnormal{in}'/\gamma_\textnormal{ex} & 0 \\
        0 & -\gamma_\textnormal{in}/\gamma_\textnormal{ex}'
    \end{pmatrix}, \nonumber \\
    U_2&=\begin{pmatrix}
        \frac{\gamma_\textnormal{in}+\gamma_\textnormal{ex}+\gamma_\textnormal{in}'+\gamma_\textnormal{ex}'}{\gamma_\textnormal{ex}} & -\frac{\gamma_\textnormal{in}+\gamma_\textnormal{ex}'}{\gamma_\textnormal{ex}} \\[0.2cm]
        -\frac{\gamma_\textnormal{in}'+\gamma_\textnormal{ex}}{\gamma_\textnormal{ex}'} & \frac{\gamma_\textnormal{in}+\gamma_\textnormal{ex}+\gamma_\textnormal{in}'+\gamma_\textnormal{ex}'}{\gamma_\textnormal{ex}'}
    \end{pmatrix}, \nonumber \\
    U_3&=\begin{pmatrix}
        -\gamma_\textnormal{ex}'/\gamma_\textnormal{in} & 0 \\
        0 & -\gamma_\textnormal{ex}/\gamma_\textnormal{in}'
    \end{pmatrix}, \nonumber \\
    U_4&=\begin{pmatrix}
        \frac{\gamma_\textnormal{in}+\gamma_\textnormal{ex}+\gamma_\textnormal{in}'+\gamma_\textnormal{ex}'}{\gamma_\textnormal{in}} & -\frac{\gamma_\textnormal{in}'+\gamma_\textnormal{ex}}{\gamma_\textnormal{in}} \\[0.2cm]
        -\frac{\gamma_\textnormal{in}+\gamma_\textnormal{ex}'}{\gamma_\textnormal{in}'} & \frac{\gamma_\textnormal{in}+\gamma_\textnormal{ex}+\gamma_\textnormal{in}'+\gamma_\textnormal{ex}'}{\gamma_\textnormal{in}'}
    \end{pmatrix}.
\end{align}
The transfer matrix has four eigenvectors corresponding to the steady state. Two of them are uniformly distributed $\bm{V}_1=[1, 1, 1, 1]^T$. One eigenvector is localized on the left edge $\bm{V}_\alpha$ with a decay constant given by the transfer matrix eigenvalue $\alpha$, while another is localized on the right with the same decay length. 
On the left edge, the following linear combination satisfies the boundary conditions:
\begin{equation}
    \bm{P}_0=\bar{p}_\textnormal{bulk}(\bm{V}_1+\xi \bm{V}_\alpha),
\end{equation}
where $\bar{p}_\textnormal{bulk}$ is the probability value in the bulk and $\xi$ is determined from the boundary conditions. Away from the left edge, the bulk portion of probability remains constant, while the edge-localized portion decays with the exponent $\alpha$. Then the steady state near the left edge is given by
\begin{equation}
    \ket{\bar{p}}=\bar{p}_\textnormal{bulk}\sum_{n=0}(\bm{V}_1 + \alpha^n\xi\bm{V}_\alpha)\otimes \ket{n}.
    \label{eq:stst}
\end{equation}
Here, $\ket{n}$ encodes a state in the $n$-th unit cell along the $x$ direction. In the main text, we analyze edge-localization in the 2D SSH model by computing the total excess of probability on the edge over the uniform bulk value 
\begin{align}
\bar{P}_\textnormal{edge}&=\sum_i(\bar{p}_i-\bar{p}_\textnormal{bulk})/\bar{p}_\textnormal{bulk}  \nonumber \\
&=\sum_{n=0}\alpha^n\xi[1, 1, 1, 1]\cdot\bm{V}_\alpha \nonumber \\
&=\frac{\xi}{1-\alpha}[1, 1, 1, 1]\cdot\bm{V}_\alpha.
\end{align}
This quantity is plotted in Fig.~\ref{fig:quasi-edge}(d).

In the steady state given by Eq.~\ref{eq:stst}, the probability flow along the left edge is computed as 
\begin{align}
    J_\textnormal{st.}&=\sum_{n=0}[\gamma_\textnormal{in},-\gamma_\textnormal{in}', \gamma_\textnormal{in}', -\gamma_\textnormal{in}]\cdot\ket{\bar{p}}_n, \nonumber\\
    &=\bar{p}_\textnormal{bulk}\frac{\xi}{1-\alpha}[\gamma_\textnormal{in},-\gamma_\textnormal{in}', \gamma_\textnormal{in}', -\gamma_\textnormal{in}]\cdot\bm{V}_\alpha.
    \label{eq:st_current}
\end{align}
The edge-localized current is plotted in Fig.~\ref{fig:edge_chirality}(g) in the units of $\bar{p}_\textnormal{bulk}\cdot\gamma_\textnormal{tot}$ as a function of non-reciprocity $c$. Stochastic edge current for the full range of ratio $r$ and non-reciprocity $c$ is plotted in Fig.~\ref{fig:current_cr}(b).

\begin{figure}
    \centering
    \includegraphics{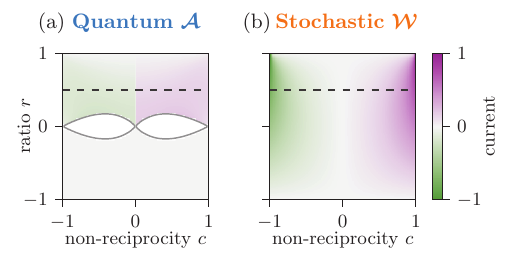}
    \caption{Persistent current flowing along the edge of the 2D SSH model as a function  of ratio $r$ and non-reciprocity $c$. (a) The quantum current is computed as the group velocity of the state with the largest imaginary energy at quarter filling. Gray lines show the phase transition from an insulating to a semimetal phase. The current is calculated only in the insulating phase. (b) The stochastic current is computed using Eq.~\ref{eq:st_current} and is shown in the units of $\bar{p}_\textnormal{bulk}\cdot\gamma_\textnormal{tot}$. The dashed line indicates parameter values at which the currents are plotted in Fig.~\ref{fig:edge_chirality}(g) in the main text.}
    \label{fig:current_cr} 
\end{figure}

\subsection{1D edge model}
\label{app:1D_edge_model}

\begin{figure*}
    \centering
    \includegraphics{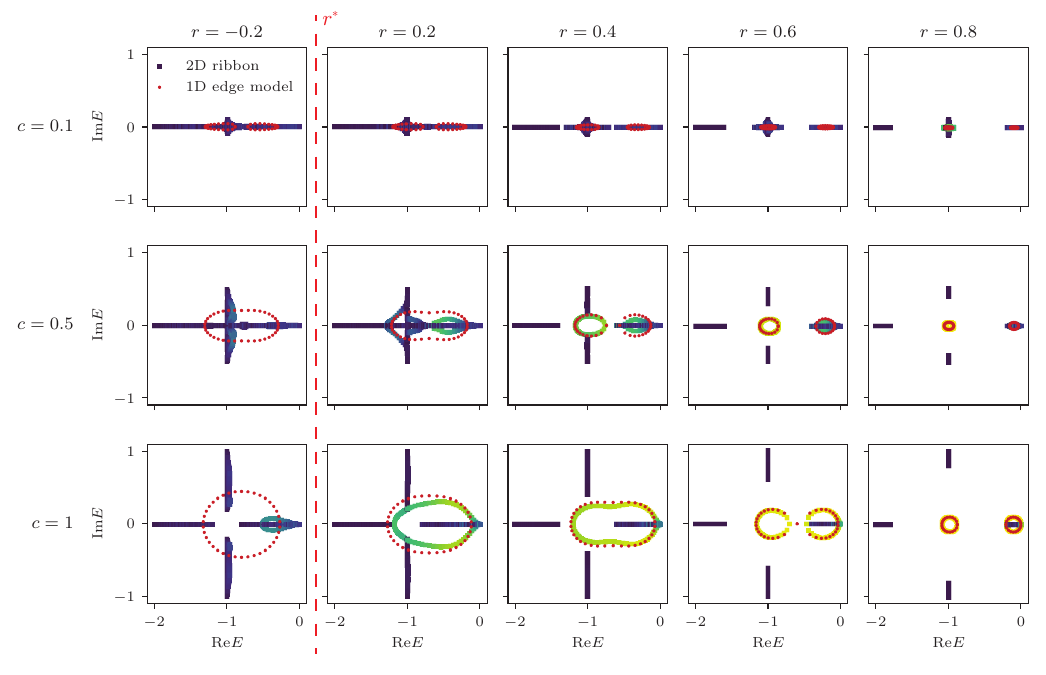}
    \caption{Comparison of the 2D SSH spectrum (purple-to-yellow squares) and the 1D edge model spectrum (red dots) for a number of model parameters $c$ and $r$. The edge-localized band of the 2D model (yellow color) can be approximated with the 1D edge model when the system is deep in the topological phase ($r>r^*$). The critical value $r^*$ with respect to the presented ratios $r$ is shown with a red dashed line.}
    \label{fig:2D_vs_edge}
\end{figure*}

In this section, we provide more details about the 1D edge model, emulating the edge of the 2D SSH model in the ribbon geometry. Its transition matrix reads
\begin{equation}
    \mathcal{W}_\textnormal{1D edge}=\begin{pmatrix}
        -\gamma_\textnormal{in}-\gamma_\textnormal{in}'-\gamma_\textnormal{ex}' & \gamma_\textnormal{in}' + \gamma_\textnormal{ex} e^{ik} \\
        \gamma_\textnormal{in} + \gamma_\textnormal{ex}' e^{-i k} & -\gamma_\textnormal{in}-\gamma_\textnormal{in}'-\gamma_\textnormal{ex}
    \end{pmatrix}.
\end{equation}
We show the similarity between the spectrum of the 1D edge model and the edge-localized band of the 2D spectrum for a number of parameters in Fig.~\ref{fig:2D_vs_edge}. Note that the 1D edge model does not conserve probability and hence does not necessarily have the steady state. We observe that deep in the topological phase when $r>r^*$, the two spectra are similar, making the 1D edge model a good approximation of the 2D edge band. Close to the phase transition and in the trivial phase $r\leq r^*$, the 1D edge model does not resemble the edge band of the 2D model. However, even in this regime, the complex spread of the 1D edge spectrum $\max\Im E/\min|\Re E|$ resembles the 2D edge localization $\bar{P}_\textnormal{edge}$ (Fig.~\ref{fig:quasi-edge}(c) vs (d)). To understand this, note that in the trivial phase the 2D localization vanishes, but so does the lifetime $(\min|\Re E|)^{-1}$ of the 1D edge model, and hence the two functions still have a similar shape.

{
\section{Kagome model}
\label{app:kagome}}

\begin{figure*}
    \centering
    \includegraphics{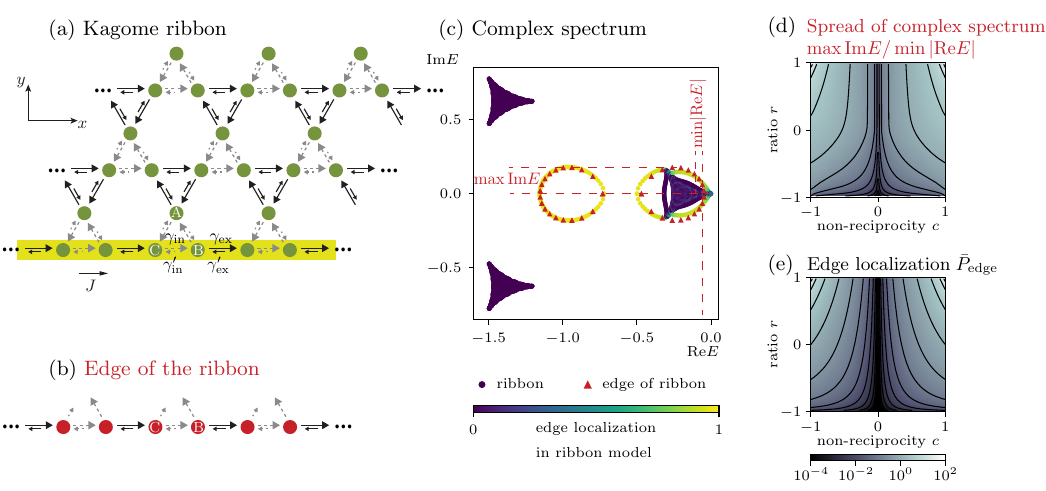}
    \caption{{\textbf{Spread of the complex spectrum induces edge localization in a stochastic kagome lattice.} (a) The kagome model in ribbon geometry has periodic boundary conditions along $x$-axis and open boundary condition along $y$-axis. The edge current is localized on the bottom edge, highlighted with yellow. (b) An auxiliary 1D model emulates the edge of the 2D kagome ribbon. (c) The spectrum of the kagome ribbon (purple vs.~yellow dots denoting bulk vs.~edge states) has edge states that are well approximated by the spectrum of the 1D edge model (red triangles). As in the 2D SSH model, we characterize the edge model spectrum by the largest imaginary component among all eigenvalues $\max\Im E$ and the smallest magnitude among all real components $\min|\Re E|$. We observe the same mechanism for edge localization in the kagome model as in the 2D SSH model. (c) The overall complex spread of the edge model, expressed as $\max\Im E/\min|\Re E|$, is similar in shape to (d) the total excess of probability on the edge of the 2D model $\bar{P}_\textnormal{edge}$.}}
    \label{fig:kagome_edge}
\end{figure*}

{In this Appendix we demonstrate that the mechanism for edge localized steady state, discussed for the 2D SSH model in Sec.~\ref{sec:edge_mechanism}, also applies to stochastic systems on a kagome lattice. This lattice, shown in  Fig.~\ref{fig:kagome_edge}(a), is constructed in the similar way as the 2D SSH model. It also contains two cycles of rates going in opposite directions, the internal and external cycles. However, its unit cell has three states, and therefore those cycles form triangles instead of rectangles. Hamiltonian in momentum space takes the form
\begin{equation}
    \mathcal{A}=\begin{pmatrix}
        0 & \gamma_\textnormal{in}+\gamma_\textnormal{ex}'e^{i\bm{k}\bm{b}_2} & \gamma_\textnormal{in}'+\gamma_\textnormal{ex}e^{-i\bm{k}\bm{b}_1}\\
        \gamma_\textnormal{in}'+\gamma_\textnormal{ex}e^{-i\bm{k}\bm{b}_2} & 0 & \gamma_\textnormal{in}+\gamma_\textnormal{ex}'e^{-ik_x}\\
        \gamma_\textnormal{in}+\gamma_\textnormal{ex}'e^{i\bm{k}\bm{b}_1} & \gamma_\textnormal{in}'+\gamma_\textnormal{ex}e^{ik_x} & 0
    \end{pmatrix}
\end{equation} 
where $\bm{k}=(k_x,k_y)$, $\bm{b}_1=(1/2, \sqrt{3}/2)$ and $\bm{b}_2=(1/2, -\sqrt{3}/2)$. The corresponding stochastic matrix is given by $\mathcal{W}=\mathcal{A}-\gamma_\textnormal{tot}I$, where $I$ is the identity matrix and $\gamma_\textnormal{tot}=\gamma_\textnormal{in}+\gamma_\textnormal{in}'+\gamma_\textnormal{ex}+\gamma_\textnormal{ex}'$.
For this model, we adopt the same parametrization as for the previously studied 1D and 2D models in terms of the ratio between external and internal rates $r$ and non-reciprocity $c$.}

{The kagome lattice has a three-fold rotation symmetry, which protects a topological invariant identified in Ref.~\cite{ni2017}. When external rates are larger than internal $r>0$, this invariant is non-zero and the system has edge currents. When $r<0$, the invariant is zero and there are no edge states. }

{Similar to the 2D SSH model, we consider the kagome model in ribbon geometry, and study currents localized at the bottom edge, highlighted with yellow in Fig.~\ref{fig:kagome_edge}(a). Spectrum of the ribbon kagome model, shown in Fig.~\ref{fig:kagome_edge}(c), has three bulk bands (purple), and an edge band (yellow). This edge band, similar to the 2D SSH model, is shifted towards the steady state by the diagonal terms $\mathcal{D}$ in the stochastic matrix. At the steady state the bulk and edge bands are mixed (green). Following Sec.~\ref{sec:edge_mechanism} in the main text, we relate the degree of steady state localization to the characteristics of the complex edge spectrum.}

{To characterize the shape of the edge band we introduce a 1D edge model that emulates the bottom edge of the 2D kagome model (see Fig.~\ref{fig:kagome_edge}(b)). We note that this edge model is exactly the same as the 1D edge model constructed from the 2D SSH model in the main text. The spectrum of the 1D edge model, shown in Fig.~\ref{fig:kagome_edge}(c) with red triangles, overlaps with the edge band of the 2D kagome model. Its shape is characterized in the same way as in the main text, as the ratio between the maximal imaginary component to the minimal magnitude of the real component $\max\textnormal{Im}E/\min|\textnormal{Re}E|$. This quantity is shown in Fig.~\ref{fig:kagome_edge}(d).}

{We compare this ratio to the edge-localization of the steady state in the full kagome model, expressed as the total excess of probability on the edge over the uniform bulk value $\bar{P}_\textnormal{edge}=\sum_i(\bar{p}_i-\bar{p}_\textnormal{bulk})/\bar{p}_\textnormal{bulk}$. We derive this quantity analytically below and plot it in Fig.~\ref{fig:kagome_edge}(e). The similarity of the quantities in Figs.~\ref{fig:kagome_edge}(d) and (e) as a function of ratio $r$ and non-reciprocity $c$ suggests that edge currents in the kagome model appear due to the spread of the complex edge spectrum. In other words, it follows the same mechanism as in the 2D SSH model, introduced in Sec.~\ref{sec:edge_mechanism}.}

{We derive the steady state probability excess on the edge using the transfer matrix formalizm introduced in Ref.~\cite{tang2021} and outlined in Appendix~\ref{app:stoch_2Dsurface}. In the kagome model in ribbon geometry, the eigenvector equation of the stochastic matrix in the $n$-th unit cell takes the form 
\begin{align*}
    (\epsilon \!+\!\gamma_\textnormal{tot})p_\mathrm{A}^n&= \gamma_\textnormal{in}p_\mathrm{B}^n+\gamma_\textnormal{ex}'p_\mathrm{B}^{n+1}e^{-\frac{ik_x}{2}}+\gamma_\textnormal{ex}p_\mathrm{C}^{n+1}+\gamma_\textnormal{in}'p_\mathrm{C}^n,\\
    (\epsilon \!+\!\gamma_\textnormal{tot})p_\mathrm{B}^n&= \gamma_\textnormal{in}'p_\mathrm{A}^n+\gamma_\textnormal{ex}p_\mathrm{A}^{n-1}e^{\frac{ik_x}{2}}\!+\gamma_\textnormal{in}p_\mathrm{C}^{n}+\gamma_\textnormal{ex}'p_\mathrm{C}^ne^{ik_x},\\
    (\epsilon \!+\!\gamma_\textnormal{tot})p_\mathrm{C}^n&= \gamma_\textnormal{in}p_\mathrm{A}^n+\gamma_\textnormal{ex}'p_\mathrm{A}^{n-1}e^{\!\!-\frac{ik_x}{2}}\!+\gamma_\textnormal{in}'p_\mathrm{B}^{n}+\gamma_\textnormal{ex}p_\mathrm{B}^ne^{\!\!-ik_x},
\end{align*}
where $\epsilon$ is the corresponding eigenvalue. To find the steady state, we set $\epsilon=0$ and $k_x=0$. We rewrite the eigenvector equations to relate the steady state probabilities in two neighboring unit cells as
\begin{align}
    \begin{pmatrix}
        p_\mathrm{A}^n\\
        p_\mathrm{B}^n\\
        p_\mathrm{C}^n
    \end{pmatrix}=U_1^{-1}U_2
    \begin{pmatrix}
        p_\mathrm{A}^{n-1}\\
        p_\mathrm{B}^{n-1}\\
        p_\mathrm{C}^{n-1}
    \end{pmatrix},
\end{align}
with the two matrices being
\begin{align}
    U_1&=\begin{pmatrix}
        0 & \gamma_\textnormal{ex}' & \gamma_\textnormal{ex} \\
        \gamma_\textnormal{in}' & -\gamma_\textnormal{tot} & \gamma_\textnormal{ex}'+\gamma_\textnormal{in} \\
        \gamma_\textnormal{in} & \gamma_\textnormal{ex}+\gamma_\textnormal{in}' & -\gamma_\textnormal{tot}
    \end{pmatrix}, \\
    U_2&=\begin{pmatrix}
        \gamma_\textnormal{tot} & -\gamma_\textnormal{in} & -\gamma_\textnormal{in}' \\
        -\gamma_\textnormal{ex} & 0 & 0 \\
        -\gamma_\textnormal{ex}' & 0 & 0
    \end{pmatrix}.
\end{align}
}

{The transfer matrix $T=U_1^{-1}U_2$ has two uniform eigenvectors $\bm{V}_b=[1, 1, 1]^T$ that correspond to bulk modes and an eigenvector $\bm{V}_e=[0,-\gamma_\textnormal{in}'/\gamma_\textnormal{in}, 1]^T$ that corresponds to an eigenvalue $\alpha=0$. This last vector gives the edge state which abruptly decays with distance from the edge. The steady state is a combination of the bulk state and the edge state, that satisfies the boundary condition on the bottom boundary
\begin{align*}
(\gamma_\textnormal{in}+\gamma_\textnormal{in}'+\gamma_\textnormal{ex})p_\mathrm{B}^0=\gamma_\textnormal{in}'p_\mathrm{A}^0+(\gamma_\textnormal{in}+\gamma_\textnormal{ex}')p_\mathrm{C}^0,\\
(\gamma_\textnormal{in}+\gamma_\textnormal{in}'+\gamma_\textnormal{ex}')p_\mathrm{C}^0=\gamma_\textnormal{in}p_\mathrm{A}^0+(\gamma_\textnormal{in}'+\gamma_\textnormal{ex})p_\mathrm{B}^0.
\end{align*}
These conditions are satisfied by the vector $\bm{P}_0=\bar{p}_\textnormal{bulk}(\bm{V}_b+\xi \bm{V}_e)$ with 
\begin{equation}
\xi=\frac{\gamma_\textnormal{in}(\gamma_\textnormal{ex}-\gamma_\textnormal{ex}')}{\gamma_\textnormal{in}^2+\gamma_\textnormal{in}'^2+\gamma_\textnormal{in}\gamma_\textnormal{in}'+\gamma_\textnormal{ex}'\gamma_\textnormal{in}+\gamma_\textnormal{in}'\gamma_\textnormal{ex}}.    
\end{equation}
This gives the probability excess on the edge 
\begin{equation}
    \bar{P}_\textnormal{edge}=\sum_i(\bar{p}_i-\bar{p}_\textnormal{bulk})/\bar{p}_\textnormal{bulk}=\xi\sum_i\bm{V}_{e,i}.
\end{equation}}

\bibliography{references}

\end{document}